# Superconductivity by Sr Intercalation in Layered Topological Insulator $Bi_2Se_3$

Shruti, Vishal K. Maurya, Prakriti Neha, Pankaj Srivastava, and Satyabrata Patnaik[*]

*School of Physical Sciences, Jawaharlal Nehru University, New Delhi-110067, India.*

ABSTRACT: Strontium intercalation between van der Waals bonded layers of topological insulator $Bi_2Se_3$ is found to induce superconductivity with a maximum $T_c$ of 2.9 K. Transport measurement on single crystal of optimally doped sample $Sr_{0.1}Bi_2Se_3$ shows weak anisotropy ($\Gamma \sim 1.5$) and upper critical field $H_{c2}(0)$ equals to 2.1 T for magnetic field applied perpendicular to $c$ -axis of the sample. The Ginzburg-Landau coherence lengths are $\xi_{ab}$ = 15.3 Å and $\xi_c$ = 10.2 Å. The lower critical field and zero temperature penetration depth $\lambda(0)$ are estimated to be 0.35 mT and 1550 nm respectively. Hall and Seebeck measurements confirm the dominance of electronic conduction and the carrier concentration is surprisingly low ($n$ = 1.85 x $10^{19}$ $cm^{-3}$) at 10 K indicating possibility of unconventional superconductivity.

Superconductors derived from topological insulating phases of quantum matter are amongst the most profound developments of recent past.[1,2,3] Extensively studied three dimensional topological insulators (TI), such as $Bi_2Se_3$ and SnTe, are characterized by gapless spin polarized surface states that emerge from the peculiar topology of insulating bulk band structure.[4] These surface states show Dirac-like linear dispersion with the spin polarization locked perpendicular to the electron momentum.[5] Correspondingly, the theoretical predictions on a superconducting analogue; a fully gapped odd parity pairing state with gapless Majorana surface states have attracted significant attention.[4] In this communication we describe synthesis and extensive electromagnetic characterization of one such example; bulk superconductivity in Sr intercalated $Bi_2Se_3$.

Pristine $Bi_2Se_3$ is a wide band semiconductor with a layered crystal structure made from double layers of $BiSe_6$ octahedra, ($R$-3m, space group 166) consisting of stacked Se–Bi–Se–Bi–Se quintuple layers that are weakly van der Waals bonded to the neighbouring sets of layers.[6] As seen in other chalcogenides,[7] this layered structure leads to intercalative chemical manoeuvrings and topological superconductivity ($T_c \sim 3.5$ K) has been reported in Cu-intercalated $Bi_2Se_3$ ($Cu_xBi_2Se_3$, 0.12 < $x$ < 0.15).[1,2] Further, at high pressure, both $Bi_2Te_3$ and $Sb_2Te_3$ are reported to be superconducting.[8,9] A recent study on In intercalated SnTe has also shown superconductivity with a $T_c$ of ~ 4.5 K.[10] Very recently, intercalation of Sr in $Bi_2Se_3$ was reported with $T_c \sim 2.5$ K with the large shielding volume fraction 88 %.[11] In this report we report successful synthesis of single crystals of Sr intercalated $Sr_xBi_2Se_3$ with an optimal $T_c$ of 2.9 K for x = 0.1 .

Single crystals of $Sr_xBi_2Se_3$ for x = 0, 0.05, 0.1, 0.15, 0.20 were prepared by melting stoichiometric ratios of high purity elements of Bi (99.999 %), Se (99.999 %) and Sr (99 %) at 850 °C in sealed evacuated quartz tubes for 8 days followed by slow cooling to 650 °C at a rate of 10 °C/h. The sample was then quenched in ice water. The obtained crystals were silvery shiny with mirror like surface and were easily cleaved along basal plane of the crystal. The samples remained silvery shiny even after exposing to air for long time unlike $Cu_xBi_2Se_3$ where the sample surface turned golden brown after one day exposure to air.[2] The dimensions of the cleaved $Sr_xBi_2Se_3$ crystals were approximately 3 x 2 x 0.5 $mm^3$. The samples were characterized by powder X-ray diffraction at room temperature using RIGAKU powder X-ray Diffractometer (Miniflex-600 with Cu-K$\alpha$ radiation). Energy Dispersive X-ray Spectroscopy (EDAX) and Scanning Electron Microscopy (SEM) images were obtained from Bruker AXS Microanalyser and from a Zeiss EVO40 SEM analyzer respectively. High Resolution Transmission Electron Microscopy (HRTEM) measurement was done using JEOL (JEM-2100F) transmission electron microscope. Resistivity, Hall and Magnetization measurements were carried out using 14 Tesla *Cryogenic* Physical Property Measurement System (PPMS). The Seebeck coefficient was measured from room temperature down to 50 K using home built set up. As discussed below, in transport and magnetization measurement it was confirmed that x = 0.1 is the optimal sample with large superconducting volume fraction and in this communication we lay stress on this composition.

Figure 1(a) shows the x-ray diffraction data along the basal plane reflections from $Sr_{0.1}Bi_2Se_3$. The sharp peak along (00l) confirms single crystalline growth of the sample. All the samples were subjected to the EDAX and the pattern for $Sr_{0.1}Bi_2Se_3$ is shown in the inset (i) of Figure 1a. Quantitative analysis of the EDAX acquired through several points on sample surface gave the percentage of the element close to the stoichiometric values. For example, for $Sr_{0.1}Bi_2Se_3$, the average atomic percentage of Sr, Bi and Se are found to be 1%, 38% and 61% respectively. All the samples show characteris-tic layered morphology as seen in inset (ii). The schematic unit cell of $Sr_{0.1}Bi_2Se_3$ is also included.

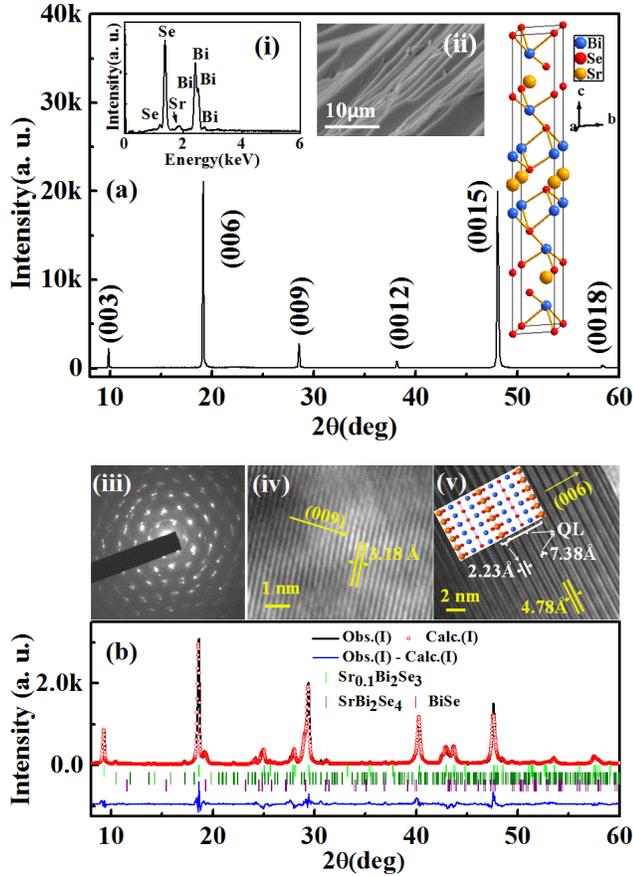

Figure 1.(a) X-ray diffraction for single crystal of $Sr_{0.1}Bi_2Se_3$. Inset (i) shows the EDAX data and Inset (ii) shows SEM image that confirms layered morphology. Schematic unit cell is also included. The main panel of Figure 1.(b) shows refined XRD patterns for $Sr_{0.1}Bi_2Se_3$. The Bragg position for impurity phases $SrBi_2Se_4$ and BiSe are marked. Inset (iii) shows diffraction pattern of $Sr_{0.1}Bi_2Se_3$. In inset (iv) the HRTEM images show (009) planes. HRTEM image of inset (v) shows (006) planes.

XRD data of $Sr_xBi_2Se_3$ samples (x = 0, 0.05, 0.1, 0.15 and 0.2) were Rietveld refined by FullProf software using reported Wyckoff positions.[12] Main panel of Figure 1b shows the refined XRD patterns of $Sr_{0.1}Bi_2Se_3$. The $Bi_2Se_3$ system crystallizes in rhombohedral structure with space group R-3m. It is understood that the Sr atom in the van der Waals gap goes to octahedrally coordinated 3b (0, 0, 1/2) sites.[2,14] Two impurity phases, BiSe and $SrBi_2Se_4$, are also present in the samples and their concentration grows with increase in Sr percentage. These impurity phases have been taken into account for better refinement. The refined lattice constants of $Bi_2Se_3$ are a = 4.14 Å and c = 28.64 Å that are in agreement with reported data.[11] Evidently, the c-axis increases after Sr intercalation with a = 4.14 Å and c = 28.65 Å for $Sr_{0.1}Bi_2Se_3$[11, 13]. The overall volume of the unit cell was found to increase from 425.22 Å to 425.38 Å, 425.40 Å, 426.20 Å, 426.55 Å for x = 0.0, 0.05, 0.1, 0.15, 0.2 respectively. This confirms successful intercalation of Sr into $Bi_2Se_3$ for 0.05 < x <0.2 .

Super-lattice pattern in electron diffraction measurements is a strong indicator of intercalation.[15] Inset (iii) in Figure 1b shows the diffraction pattern for $Sr_{0.1}Bi_2Se_3$ which demonstrates diffused, azimuthal spread of the superlattice satellite spots associated with each $Bi_2Se_3$ Bragg spot. The diffuse nature of spots suggest lower intercalant concentration and the azimuthal spread is due to large atomic radii of Sr.[15] Inset (iv) and Inset (v) show the HRTEM images for $Sr_{0.1}Bi_2Se_3$ along (009) and (006) plane respectively. The HRTEM shows no sign of stacking faults, intergrowths or amorphous regions. The d values measured from the fringes for $Sr_{0.1}Bi_2Se_3$ corresponding to (009) and (006) planes are 3.18 Å and 4.78 Å respectively. In inset (v) the width of the dark lines corresponds to distance between the quintuple layers. Moreover, there is a small and broad gap between fringes that appear alternatively. The broader gap corresponds to the spacing between two quintuple layers (QL) as shown in schematic for (006) plane (Figure 1b (v). The measured spacing of 2.23 Å from HRTEM image corroborate well with XRD result.

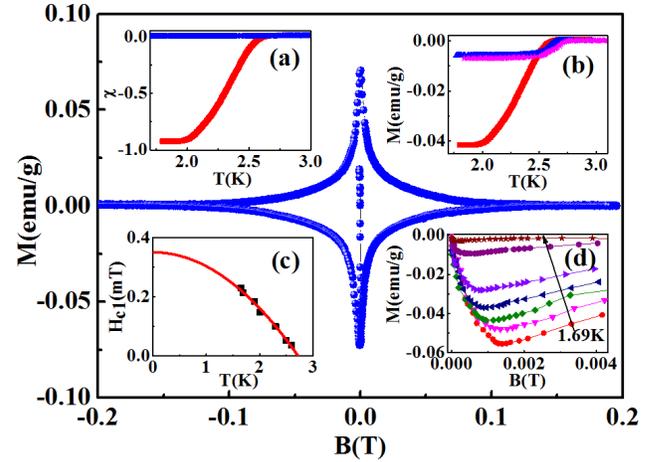

Figure 2. The M-H loop with H||ab plane at 1.65 K is shown for $Sr_{0.1}Bi_2Se_3$. In inset (a) the DC susceptibility measured with H||ab plane under ZFC (red) and (FC) (blue) condition are plotted for 5G field. Inset (b) shows the DC magnetization graphs under ZFC for x = 0.1, 0.15, and 0.2 Sr intercalated $Bi_2Se_3$ samples. Inset (c) shows lower critical field $H_{c1}$ data points (black) and parabolic fitting (red) line. Inset (d) shows M-H scan in low field range at T= 1.69, 1.9, 1.97, 2.0, 2.3, 2.5 and 2.6 K.

The DC magnetization measured for $Sr_{0.1}Bi_2Se_3$ with H||ab under Zero field cooled (ZFC) and Field cooled (FC) protocols using 5 Gauss magnetic field is shown in inset (a) of Figure 2. Onset of diamagnetic signal is marked at 2.67 K. The superconducting shielding fraction is about 93 % which shows the bulk nature of superconductivity in this compound. The magnetization data taken in FC measurement does not show significant flux trapping implying the absence of effective pinning centres in the single crystals. Inset (b) of Figure 2 shows same measurement for x = 0.1, 0.15, and 0.2 $Sr_xBi_2Se_3$ samples. All the samples within the composition range 0.1 < x < 0.2 showed superconductivity with a maximum of 2.75 K for x = 0.2, but the superconducting volume fraction decreased with increase in Sr concentration. The M-H loop with H||ab at 1.65 K shown in main panel of Figure 2 reflects type-II supercon-

ductivity with identifiable irreversibility and upper critical fields. The inset (d) in Figure 2 shows the M-H scan in smaller field range at different temperatures towards measuring the lower critical field which is marked by deviation from linearity in the diamagnetic state. Using this criterion, a set of $H_{c1}(T)$ was estimated and fitted to the parabolic T dependence $H_{c1}(T) = H_{c1}(0)[1-(T/T_c)^2]$, $H_{c1}(0)$ is estimated to be 3.5 ± 0.1 G. This is shown in inset (c) of Figure 2.

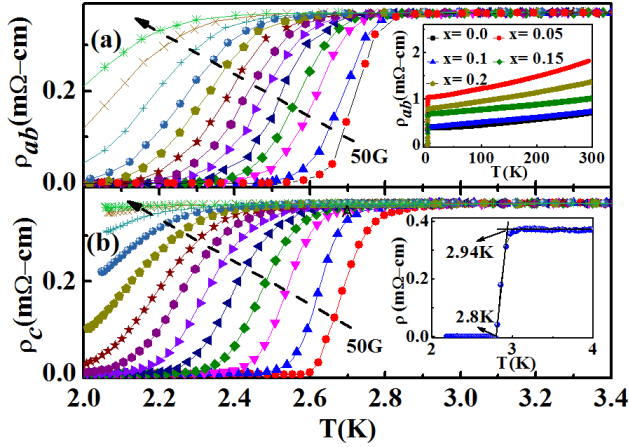

Figure 3. The resistivity as a function of magnetic field applied parallel to *ab* plane (a) and parallel to *c*-axis (b) is plotted for $Sr_{0.1}Bi_2Se_3$. External field H = 0.005, 0.02, 0.05, 0.08, 0.11, 0.14, 0.17, 0.20, 0.24, 0.28, 0.34, 0.42 and 0.50 T. Inset in (a) shows the resistivity behaviour from 2 K to 300 K. Inset in (b) shows the expanded view of the superconducting transition.

In Figure 3 we summarize magneto-transport measurements. Inset in Figure 3a shows the resistivity from low temperature to room temperature of $Sr_xBi_2Se_3$. For $Sr_{0.1}Bi_2Se_3$ we find a $T_{c-onset}$ 2.94 K and $T_{c-zero}$ 2.8 K. This is higher by ~ 0.35 K as compared to previous report.[11] Further increase of Sr concentration does not significantly affect resistive $T_c$ but the impurity concentration goes up. The resistivity of $Sr_{0.1}Bi_2Se_3$ is linear in entire temperature range with residual resistivity $\rho_0$ = 0.37 x $10^{-5}$ Ω-m. A sharp superconducting transition with transition width $\Delta T_c = T_{c-onset} - T_{c-zero}$ = 0.16 K (inset of Figure 3b) and residual resistivity ratio ($\rho_{300K}/\rho_{3K}$) = 1.95 are observed. As shown in the main panel of Figure 3, with the application of external magnetic field, a larger shift in superconducting transition is observed for $H||c$- axis than $H||ab$ plane. This behaviour is reflective of anisotropic transport in layered superconductors as evidenced in $Cu_xBi_2Se_3$ as well.[2]

From the onset and offset data of superconducting transition in the presence of magnetic field (Figure 3), in the inset of Figure 4 we plot the upper critical field $H_{c2}(T)$ and Irreversibility field ($H_{irr}$) as a function of temperature. The standard criterion used for these estimations is described elsewhere.[20] Further, in a weak coupling case, the Pauli limited upper critical field is given by $H_P(0) = 1.84 \times T_c$ = 5.15 T. Under generalized Gingburg Landau model; $H_{c2}(T) = H_{c2}(0)[(1-t)^2/(1+t^2)]$ where $t = T/T_c$. In the main panel of Figure 4 fitting of our data to this equation gives $H_{c2}||ab(0)$ = 2.1 ± 0.025 T (red line) and $H_{c2}||c(0)$ = 1.4 ± 0.035 T (blue line). In conventional single band Werthamer Helfand - Hohenberg (WHH) theory,[16] the orbital limited upper critical field of type – II superconductor in the clean limit is be described by $H_{c2}^{orb}(0) = 0.72 \times T_c \times (dH_{c2}/dT|_{Tc})$ and estimated to be 1.6 T. Since $H_{c2}^{orb}(0) < H_{c2}(0) < H_P(0)$ the superconductivity in $Sr_{0.1}Bi_2Se_3$ is Pauli limited. With Maki parameter $\alpha = \sqrt{2}H_{c2}^{orb}(0)/H_P(0)$ less than 1, the Flude - Ferrell – Larkin - Ovchinnikov (FFLO) state, which is reflective of spatially modulated order parameter, is not indicated.[17] The electronic anisotropy parameter $\Gamma = H_{c2,ab}/H_{c2,c}$ is about 1.5 and using the anisotropic Ginzburg-Landau formulas[18] $H_{c2}||^{ab} = \Phi_0/(2\pi \xi_{ab} \xi_c)$ and $H_{c2}||^c = \Phi_0/(2\pi \xi_{ab}^2)$, with $\Phi_0$ is the flux quantum $\Phi_0$ = 2.07 x $10^{-7}$ G-$cm^2$, the Ginzburg-Landau (GL) coherence lengths are estimated as $\xi_{ab}$ = 15.3 nm and $\xi_c$ = 10.2 nm. The London penetration depth at zero temperature $\lambda(0)$ is related to $H_{c1}(0)$ by the equation[19] $H_{c1} = (\Phi_0/4\pi\lambda^2)[\ln(\lambda/\xi)+0.485]$ and gives $\lambda(0)$ ~ 1550 nm. The Ginzburg Landau parameter $\kappa = \lambda/\xi$ = 101.

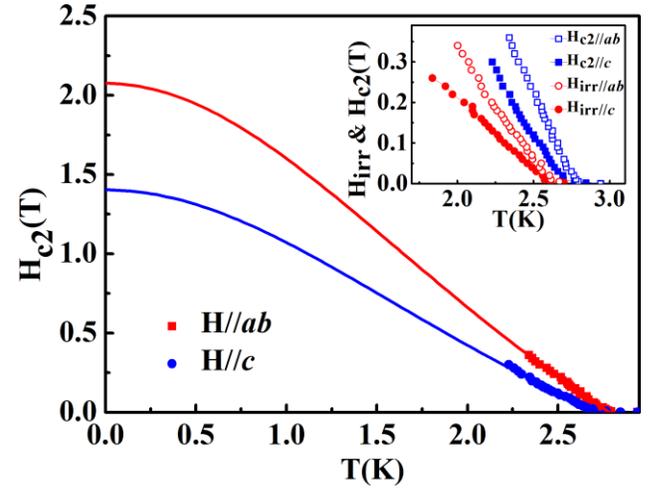

Figure 4. The extrapolated Ginzburg-Landau upper critical field ($H_{c2}$) as a function of temperature for superconducting $Sr_{0.1}Bi_2Se_3$. Inset shows the upper critical field $H_{c2}(T)$ and $H_{irr}(T)$ found from onset and offset temperature for magnetic field applied parallel and perpendicular to *c*-axis.

To shed more light on the normal state transport properties of the compound in Figure 5 we present Hall and Thermoelectric Power data on single crystal of $Sr_{0.1}Bi_2Se_3$. As shown in inset (a) of Figure 5, the Seebeck coefficient S remains negative at all temperatures, indicating electronic charge transport. Inset (b) of Figure 5 shows the magnetic field dependence of transverse resistivity at 10 K. The $\rho_{xy}$ shows linear behaviour with magnetic field and is negative. The main panel of Figure 5 shows the Hall coefficient $R_H$ as a function of temperature. $R_H$ is found to be increasing monotonically with temperature up to 200 K. From $R_H$ (= -1/ $n$e) charge carrier density $n$ has been plotted with respect to temperature in the inset (c) of Figure 5. The value of $n$ = 1.85 x $10^{19}$ $cm^{-3}$ at 10 K is one order of magnitude less than the Cu intercalated $Bi_2Se_3$.[2] It is indeed surprising to see superconductivity at 2.9 K with such low carrier concentration and most likely a non-BCS mechanism is at play as is the case in $Cu_xBi_2e_3$.[25] The normal state magneto-resistance (not shown) was found to be non- linear for field upto 12 Tesla.

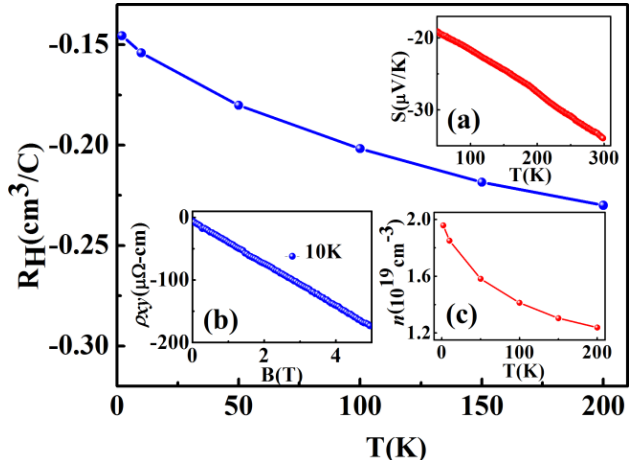

Figure 5. The Hall coefficient $R_H$ as a function of temperature measured on single crystal of $Sr_{0.1}Bi_2Se_3$ is plotted. Inset (a) shows the temperature dependence of Seebeck coefficient (S). Inset (b) shows the magnetic field dependence of transverse resistivity at 10 K. Charge carrier density $n$ is plotted as a function of temperature in inset (c).

The low temperature Hall measurement in combination with TEP data can be used to estimate effective mass of carriers $m^*$ and Sommerfeld constant $\gamma$.[20] Using simplistic free electron model and with electron carrier concentration $1.58 \times 10^{25}$ m$^{-3}$ at 50 K, the Fermi energy (Temperature) comes out to be 23 meV (266.8 K). In low carrier concentration metals, the Seebeck coefficient is given by the Mott expression $S = \pi^2 k_B T/(2eT_F)$, where $T_F$ is the Fermi temperature.[21,22] Using $T_F$ = 266.8 K in this expression yields Seebeck coefficient of -80 $\mu$VK$^{-1}$ at 50 K which is larger than the measured value of -19 $\mu$VK$^{-1}$. This implies effective mass of electron $m^* = 0.24 m_e$ and corresponding value of Sommerfeld constant $\gamma$ equals to 4.0 J/m$^3$K$^2$. The value of $m^*$ is 1.85 times higher than its value in parent $Bi_2Se_3$ as calculated from ARPES and SdH measurement.[23,24] We note that the mean free path for charge carriers are given by the relation $l = hk_F/2\pi\rho_0 ne$.[2,25] Taking $\rho_0 = 0.37 \times 10^{-5}$ $\Omega$-m and $n = 1.85 \times 10^{25}$ m$^{-3}$ at 10 K and assuming a spherical Fermi surface with wave vector $k_F = (3\pi^2 n)^{1/3} = 8.2 \times 10^8$ m$^{-1}$, the mean free path $l$ is estimated to be 49.4 nm which is greater than $\xi_{ab}$ and $\xi_c$. This confirms that our samples are in clean limits.

In summary, we report superconductivity in Sr intercalated topological insulator $Bi_2Se_3$ with Sr concentration in the range $0.1 < x < 0.2$. This is achieved at a surprisingly low carrier concentration of $1.85 \times 10^{19}$ cm$^{-3}$. About 93 % superconducting volume fraction is obtained for $x = 0.1$ with onset resistive $T_c = 2.94$ K. The anisotropy in upper critical field of $Sr_{0.1}Bi_2Se_3$ yields $\Gamma = 1.5$ with upper critical field $H_{c2}(0)$ equal to 2.1 T for magnetic field applied parallel to $ab$ plane of the sample. Corresponding Ginzburg-Landau coherence lengths are $\xi_{ab}$ = 15.3 Å and $\xi_c$ = 10.2 Å. The lower critical field and zero temperature penetration depth are 0.35 mT and 1550 nm respectively. Signatures of unconventional pairing state are indicated.

## AUTHOR INFORMATION

Corresponding Author

spatnaik@mail.jnu.ac.in

Notes

The authors declare no competing financial interests.

## ACKNOWLEDGMENT

Shruti, and V.K.M. acknowledge U.G.C. India, for fellowships. P.S. acknowledges U.G.C. for D.S. Kothari fellowship. SP thanks FIST program of DST India and Technical Support from AIRF (JNU).